\documentclass[letterpaper, 10 pt, conference]{ieeeconf}  

\IEEEoverridecommandlockouts                              
\usepackage{amsmath,amssymb}
\usepackage{mathtools}
\usepackage{cite}
\usepackage{ntheorem}
\usepackage{subcaption}
\usepackage{booktabs}
\usepackage{paralist}
\usepackage{xcolor}
\usepackage[hidelinks]{hyperref}

\usepackage{caption}
\usepackage[noabbrev]{cleveref}

\crefname{equation}{}{}

\newtheorem{remark}{Remark}
\newtheorem{proposition}{Proposition}
\newtheorem{assumption}{Assumption}
\newtheorem{theorem}{Theorem}

\title{\LARGE \bf
Robust Output Feedback MPC with \\ Reduced Conservatism under Ellipsoidal Uncertainty
}

\author{Tianchen Ji$^{1}$, Junyi Geng$^{2}$, and Katherine Driggs-Campbell$^{1}$
\thanks{$^{1}$T. Ji and K. Driggs-Campbell are with the Department of Electrical and Computer Engineering at the University of Illinois at Urbana-Champaign.
        emails: {\tt\small \{tj12,krdc\}@illinois.edu}}%
\thanks{$^{2}$Junyi Geng is with the Robotics Institute at the Carnegie Mellon University.
        email: {\tt\small junyigen@andrew.cmu.edu}}%
  \thanks{This material is based upon work supported by the National Science Foundation under Grant No. 2143435.}%
}

\begin{document}

\maketitle
\thispagestyle{empty}
\pagestyle{empty}

\begin{abstract}

Robust design of autonomous systems under uncertainty is an important yet challenging problem. This work proposes a robust controller that consists of a state estimator and a tube based predictive control law. The class of linear systems under ellipsoidal uncertainty is considered. In contrast to existing approaches based on polytopic sets, the constraint tightening is directly computed from the ellipsoidal sets of disturbances without over-approximation, thus leading to less conservative bounds. Conditions to guarantee robust constraint satisfaction and robust stability are presented. Further, by avoiding the usage of Minkowski sum in set computation, the proposed approach can also scale up to high-dimensional systems. The results are illustrated by examples.
\end{abstract}

\section{Introduction}
Model predictive control (MPC) is a control technique based on the iterative solution of an optimization problem~\cite{bemporad1999robust}. By using the system model and the current state, MPC plans the optimal control sequence based on a cost function. The system executes the first control input in the optimal sequence and the procedure repeats at the next time step. MPC has received considerable attention over the last decades driven largely by its ability to handle multi-variable systems and state/input constraints~\cite{qiu2019output}.

In practice, however, two important issues arise in MPC design:
\begin{inparaenum}[(\itshape i)]
\item
the actual state is often not available, leading to the necessity of state estimation, and
\item
measurements and the model used for prediction are uncertain (e.g., due to disturbances and unmodeled dynamics).
\end{inparaenum}
Therefore, robust output feedback MPC schemes have been proposed and investigated to overcome the above challenges, see~\cite{findeisen2003state,mayne2006robust,mayne2009robust,kogel2017robust,de2020robust}.

Tube based output feedback MPC uses a combination of a state estimator with a robust model predictive control law~\cite{mayne2006robust,mayne2009robust,kogel2017robust,kogel2021robust,lorenzetti2020simple}. The basic idea is to decompose the closed-loop dynamics into a nominal, disturbance free system for prediction and optimization and utilizes one or two tubes to handle the uncertainties. Other robust output feedback MPC approaches are based on min-max optimization~\cite{jia2005min,findeisen2004min}, moving horizon estimation~\cite{sui2008robust,kogel2015robustreduced}, set-membership estimation~\cite{bemporad2000output,chisci2002feasibility,qiu2019output,dong2020homothetic}, or combinations of linear and set-membership estimation~\cite{brunner2018enhancing,kogel2015robusttriggered}.

\begin{figure}[t]
  \centering
  \begin{subfigure}[b]{0.46\linewidth}
    \includegraphics[width=\linewidth]{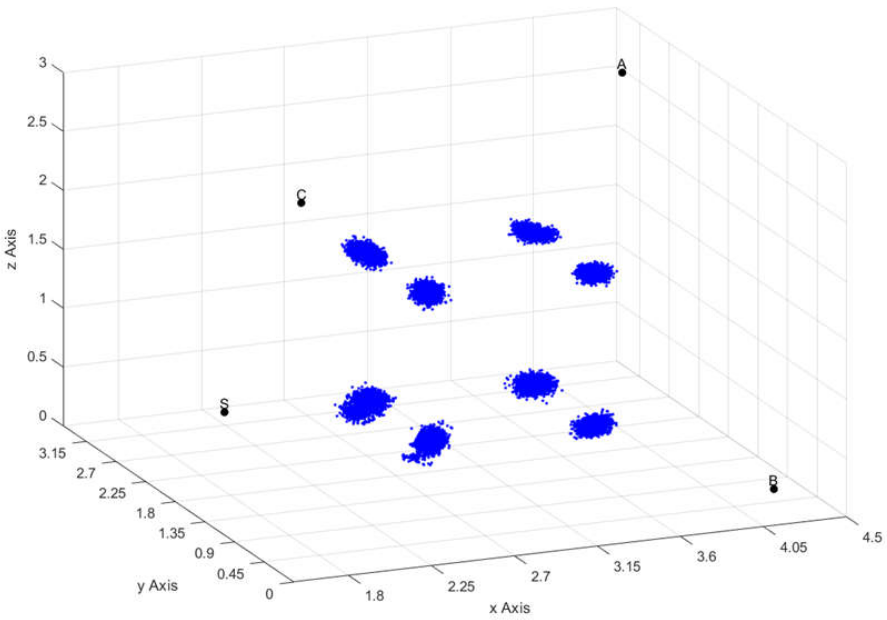}
  \end{subfigure}
  \begin{subfigure}[b]{0.5\linewidth}
    \includegraphics[width=\linewidth]{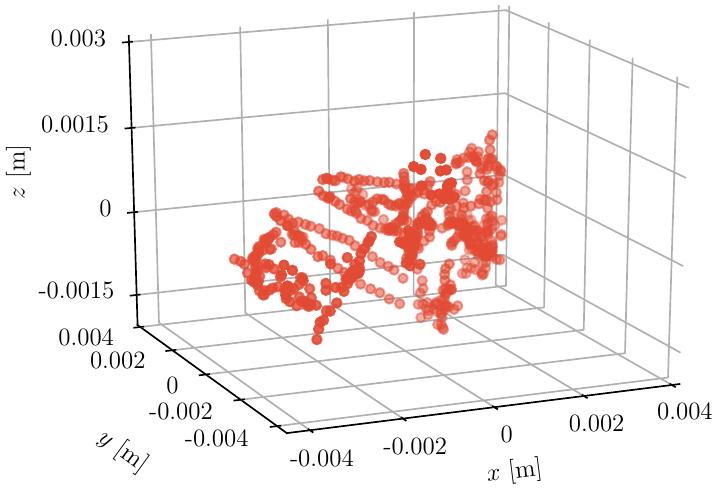}
  \end{subfigure}
  \caption{Real disturbances in different systems turn out to lie in ellipsoidal bounding sets. \textit{Left}~\cite{decawave2021measurements}: Measurements for eight different test points from Decawave indoor positioning system. Blue dots indicate the measurement results and black dots indicate the places of anchors. \textit{Right}: RTK GPS position measurements on a static Polaris GEM autonomous vehicle~\cite{du2020online}. The measurements have been normalized to have zero mean.}
  \label{fig.intro}
  \vspace{-5mm}
\end{figure}

In problem settings of robust output feedback MPC, bounding sets for disturbances are often described by polytopes~\cite{mayne2006robust,mayne2009robust,kogel2017robust,de2020robust,le2011robust}. In real-world experiments, however, we observe that ellipsoids turn out to be a more suitable description of disturbance sets, as shown in Figure~\ref{fig.intro} and in~\cite{feng2018ellipsoidal,manchester2017dirtrel}. To apply the previous approaches under such uncertainties, over-approximation of ellipsoids by polytopes needs to be performed, which is in general conservative. This work focuses on an extension of the tube-based output feedback MPC~\cite{mayne2006robust,mayne2009robust,kogel2017robust} to ellipsoidal uncertainty. The main contribution is a robust controller that can achieve less conservative results and can scale up to high-dimensional systems. To this end, we derive the constraint tightening directly from ellipsoidal sets, which allows to obtain tighter approximations on the worst case effect of the uncertainties. Moreover, our approach avoids the usage of Minkowski sum in constraint tightening and thus is more computationally efficient in high-dimensional systems.

\textit{Nomenclature}: For two sets $\mathcal{U}$, $\mathcal{V}$, $\mathcal{U} \oplus \mathcal{V}$ and $\mathcal{U} \times \mathcal{V}$ denote the Minkowski sum and Cartesian product. A set $\mathcal{U}$ is a $C$ set if it is compact, convex, and contains the origin. 
\section{Problem Setup}
We consider the following linear, uncertain, discrete-time, time-invariant system:
\begin{subequations}
\label{eq:system}
\begin{align}
\label{eq:system-x}
x_{k+1} &= Ax_k + Bu_k + w_k, \\
\label{eq:system-y}
y_k &= Cx_k + v_k,
\end{align}
\end{subequations}
where $x_k \in \mathbb{R}^n$ is the system state at time $k$, $u_k \in \mathbb{R}^m$ is the control input, $y_k \in \mathbb{R}^p$ is the measured output, $w_k \in \mathbb{R}^q$ is an unknown state disturbance, $v_k \in \mathbb{R}^p$ is an unknown output disturbance, and $(A,B,C)$ are known matrices, where the couple $(A,B)$ is assumed to be controllable and $(A,C)$ observable. The state disturbance $w_k$ and the output disturbance $v_k$ are only known to the extent that they lie, respectively, in the \textit{ellipsoidal} $C$ sets $\mathbb{W}$ and $\mathbb{V}$:
\begin{align}
\label{eq:disturbances}
\begin{split}
w_k \in \mathbb{W} &\coloneqq \{w \mid w^\top Q^{-1} w \le 1\}, \\
v_k \in \mathbb{V} &\coloneqq \{v \mid v^\top R^{-1} v \le 1\},
\end{split}
\end{align}
where $Q$ and $R$ are known positive definite matrices.

For the initial state $x_0$, an estimate $\hat{x}_0$ is available satisfying:
\begin{equation}
\label{eq:initial-uncertainty}
x_0 - \hat{x}_0 \in \mathbb{E}_0 \coloneqq \{e \mid e^\top \Psi^{-1} e \le 1\},
\end{equation}
where $\mathbb{E}_0$, the initial uncertainty, is an \textit{ellipsoidal} $C$ set.

System~(\ref{eq:system}) is subject to the following mixed constraints on the system state and control input:
\begin{equation}
\label{ineq:constraints}
Fx_k + Gu_k \le f,
\end{equation}
where $(F,G,f) \in \mathbb{R}^{d\times n} \times \mathbb{R}^{d \times m} \times \mathbb{R}^d$ are known matrices. Note that the \textit{ellipsoidal} uncertainty~\cref{eq:disturbances,eq:initial-uncertainty} and the \textit{polytopic} constraint~(\ref{ineq:constraints}) are considered.

The overall goal is to design a robust output feedback controller for system~(\ref{eq:system}) such that the closed-loop system is robustly stabilized and the constraint~(\ref{ineq:constraints}) is satisfied for any admissible state disturbance sequence $\mathbf{w} \coloneqq \{w_0,w_1,\dots\}$, output disturbance sequence $\mathbf{v} \coloneqq \{v_0,v_1,\dots\}$, initial uncertainty $x_0 - \hat{x}_0$, and $k \ge 0$. The proposed controller is based on the combination of a state estimator with a nonlinear feedback law based on MPC as in~\cite{mayne2006robust,mayne2009robust,kogel2017robust,kogel2021robust}.
\section{Bounding the Uncertainty}
\label{sec:cstr-tightening}
In this section, we review previous methods using \textit{polytopic} sets and propose a novel approach using \textit{ellipsoidal} sets to tighten the constraint to account for the uncertainty in the closed-loop system.

\subsection{Uncertainty bounding via polytopic sets}
The previous works in tube-based output feedback MPC utilizes a simple Luenberger observer to estimate the state. The constraint tightening is achieved by bounding the estimation error and the prediction error using polytopes.

\subsubsection{State estimate}
A linear observer of the form
\begin{equation*}
\hat{x}_{k+1} = A \hat{x}_k + Bu_k + L(C\hat{x}_k - y_k)
\end{equation*}
is used, where the observer gain $L$ needs to be chosen such that $A + LC$ has eigenvalues only inside the unit disc.

\subsubsection{State decomposition}
The state estimate $\hat{x}_k$ is used to calculate the control $u_k$. Therefore, the previous works decompose the state $x_k$ into three different components, compare~\cite{mayne2006robust,mayne2009robust,kogel2017robust,kogel2021robust,lorenzetti2020simple}. The first component is the nominal state $\bar{x}_k$, whose dynamics depends on the nominal input $\bar{u}_k$:
\begin{equation}
\label{eq:nominal-system}
\bar{x}_{k+1} = A\bar{x}_k + B\bar{u}_k.
\end{equation}
The second component is the estimation error $e_k$, which is the difference between the real state $x_k$ and the state estimate $\hat{x}_k$:
\begin{equation}
\label{eq:e}
e_k = x_k - \hat{x}_k.
\end{equation}
The last component is the prediction error $\xi_k$, which is the difference between the state estimate $\hat{x}_k$ and the nominal state $\bar{x}_k$:
\begin{equation*}
\xi_k = \hat{x}_k - \bar{x}_k.
\end{equation*}
The real state $x_k$ can then be represented by:
\begin{equation}
\label{eq:state-decomposition-three}
x_k = \bar{x}_k + e_k + \xi_k.
\end{equation}
The decomposition allows to design a predictive control law for the nominal system~(\ref{eq:nominal-system}), which is unaffected by the noise, and to derive the constraint tightening for the satisfaction of~(\ref{ineq:constraints}) by considering the uncertainty in $e_k$ and $\xi_k$.

\subsubsection{Control input calculation}
A control law combining a feed-forward component, given by the tube-based model predictive controller, and a feedback component is considered:
\begin{equation}
\label{eq:control-input}
u_k = \bar{u}_k + K(\hat{x}_k - \bar{x}_k),
\end{equation}
where $K$ is a fixed feedback gain satisfying that $A + BK$ has eigenvalues only inside the unit disc.

\subsubsection{Constraint tightening via separate sets}
The above definitions give rise to the following dynamics for the estimation error $e_k$:
\begin{equation}
\label{eq:separate-sets-e}
e_{k+1} = (A + LC)e_k + w_k + Lv_k.
\end{equation}
The prediction error $\xi_k$ evolves according to:
\begin{equation}
\label{eq:separate-sets-xi}
\xi_{k+1} = (A + BK)\xi_k - L(Ce_k + v_k).
\end{equation}
With the initial uncertainty $\mathbb{E}_0$, the bounds on the estimation error $e_k$ is governed by:
\begin{equation}
\label{eq:separate-sets-recursion-E}
\mathbb{E}_{k+1} = (A + LC)\mathbb{E}_k \oplus \mathbb{W} \oplus L\mathbb{V}.
\end{equation}
To bound $\xi_k$, Mayne et al.~\cite{mayne2006robust,mayne2009robust} treated the observer correction term $-L(Ce_k + v_k)$ in~(\ref{eq:separate-sets-xi}) as an ``artificial" disturbance $\phi_k$ satisfying:
\begin{equation*}
\phi_k \in \Phi_k, \quad \Phi_k = -LC\mathbb{E}_k \oplus -L\mathbb{V},
\end{equation*}
which results in the set recursion:
\begin{equation}
\label{eq:separate-sets-recursion-Xi}
\Xi_{k+1} = (A + BK)\Xi_k \oplus \Phi_k.
\end{equation}
To achieve less conservative constraint tightening, $\Xi_0=\{0\}$ can be used by setting $\bar{x}_0=\hat{x}_0$, compare~\cite{lorenzetti2020simple,kogel2021robust}. As a result, the constraint~(\ref{ineq:constraints}) can be robustly satisfied if the following inequality holds for any $\xi_k \in \Xi_k$, $e_k \in \mathbb{E}_k$, and $k \ge 0$:
\begin{equation}
\label{ineq:cstr-tightening-separate-sets}
F\bar{x}_k + G\bar{u}_k + (F + GK)\xi_k + Fe_k \le f.
\end{equation}
It has been shown in~\cite{mayne2009robust} that the sets $\mathbb{E}_k$ and $\Xi_k$ converge in the Hausdorff metric to the minimal robust positive invariant (RPI) sets $\mathbb{E}_\infty$ and $\Xi_\infty$, respectively, satisfying:
\begin{align*}
\begin{split}
\mathbb{E}_\infty &= (A + LC)\mathbb{E}_\infty \oplus \mathbb{W} \oplus L\mathbb{V}, \\
\Xi_\infty &= (A + BK)\Xi_\infty \oplus -LC \mathbb{E}_\infty \oplus -L \mathbb{V}.
\end{split}
\end{align*}
The above minimal RPI sets allow to compute only a finite number of sets $\mathbb{E}_k$ and $\Xi_k$ and to over-approximate the sets consistently after a specific $k$ for the constraint tightening, see~\cite{kogel2017robust,kogel2021robust}.

\subsubsection{Constraint tightening via a single set}
K{\"o}gel et al.~\cite{kogel2017robust,kogel2021robust} bounds the errors $e_k$ and $\xi_k$ using a single set, instead of using two separate coupled sets as discussed above. To this end, the two errors~(\ref{eq:separate-sets-e}) and~(\ref{eq:separate-sets-xi}) are combined into a composite system using the state
$
z_k
=
\begin{pmatrix}
e_k^\top & \xi_k^\top
\end{pmatrix}^\top
$:
\begin{equation*}
z_{k+1} = \tilde{A} z_k + \tilde{B}d_k, \quad d_k \in \mathbb{D}_k,
\end{equation*}
where $\mathbb{D}_k = \mathbb{W} \times \mathbb{V}$ and
\begin{equation*}
\tilde{A}
=
\begin{pmatrix}
A + LC & 0 \\
-LC & A + BK
\end{pmatrix},
\quad
\tilde{B}
=
\begin{pmatrix}
I & L \\
0 & -L
\end{pmatrix}.
\end{equation*}
With the initial composite state $z_0 \in \mathbb{Z}_0 = \mathbb{E}_0 \times \{0\}$ as above, one can have $z_k \in \mathbb{Z}_k$ where
\begin{equation}
\label{eq:single-set-recursion-Z}
\mathbb{Z}_{k+1} = \tilde{A} \mathbb{Z}_k \oplus \tilde{B} \mathbb{D}_k.
\end{equation}
Therefore, the constraint~(\ref{ineq:constraints}) can be robustly satisfied if the following inequality holds for any $z_k \in \mathbb{Z}_k$ and $k \ge 0$:
\begin{equation}
\label{ineq:cstr-tightening-single-set}
F\bar{x}_k + G\bar{u}_k +
\begin{pmatrix}
F & F + GK
\end{pmatrix}z_k
\le f.
\end{equation}
Similarly, the sets $\mathbb{Z}_k$ converge in the Hausdorff metric to the minimal RPI set $\mathbb{Z}_\infty$ satisfying
\begin{equation*}
\mathbb{Z}_\infty = \tilde{A} \mathbb{Z}_\infty \oplus \tilde{B} \mathbb{D}_k,
\end{equation*}
which again allows to bound $\mathbb{Z}_k$ by computing only a finite number of sets~\cite{kogel2017robust}. The constraint tightening~(\ref{ineq:cstr-tightening-single-set}) is less conservative than~(\ref{ineq:cstr-tightening-separate-sets}) because, for example, the errors $e_k$ and $\xi_k$ are each influenced by $v_k$, but the sum $e_k + \xi_k$ is independent of $v_k$, compare~\cref{eq:separate-sets-e,eq:separate-sets-xi}.

\begin{remark}[Bounding set representation]{\ \\}
Both the two set approach and the single set approach calculate the error bounds $\mathbb{E}_k$, $\Xi_k$, $\mathbb{Z}_k$, $\mathbb{E}_\infty$, $\Xi_\infty$, $\mathbb{Z}_\infty$ using polytopes. To apply the methods, the ellipsoidal uncertainty $\mathbb{W}$, $\mathbb{V}$, $\mathbb{E}_0$ need to be over-approximated by polytopes, which is in general conservative. Moreover, the computation of the minimal RPI sets in form of polytopes often requires Minkowski sum, which can be computationally challenging in high-dimensional systems~\cite{kogel2017robust}.
\end{remark}

\subsection{Improved bounds using ellipsoidal sets}
In contrast to the prior work discussed above~\cite{mayne2006robust,mayne2009robust,kogel2017robust,lorenzetti2020simple,kogel2021robust}, we derive the constraint tightening directly from ellipsoidal sets, which avoids the conservatism introduced by the over-approximation of the ellipsoidal uncertainty by polytopes. Furthermore, we decompose the state $x_t$ into two components rather than three components and bound a so-called control error for the constraint tightening.

\subsubsection{State estimate}
To estimate the state a classic set-membership state estimation algorithm is employed and is outlined as follows~\cite{bertsekas1971recursive}. Given the system~(\ref{eq:system}), the disturbance uncertainty~(\ref{eq:disturbances}), and the initial uncertainty~(\ref{eq:initial-uncertainty}), a bounding set $X_{k|k}$ to the set of all possible states $x_k$ at time $k$ given the outputs observed up to time $k$ can be described as an ellipsoid:
\begin{equation}
\label{eq.sse_1}
x_k \in X_{k|k} = \{x \mid (x - \hat{x}_k)^\top P_{k|k}^{-1} (x - \hat{x}_k) \le 1 - \delta_k^2\},
\end{equation}
where the positive definite matrix $P_{k|k}$ is recursively given by the equations:
\begin{align}
\label{eq.sse_2}
\begin{split}
P_{k+1|k+1} &= [(1-\rho) P_{k+1|k}^{-1} + \rho C^\top R^{-1}C]^{-1}, \\
P_{k+1|k} &= (1-\beta)^{-1}AP_{k|k}A^\top + \beta^{-1}Q, \\
P_{0|0} &= \Psi.
\end{split}
\end{align}
The estimate $\hat{x}_k$ evolves according to:
\begin{align}
\label{eq.sse_3}
\begin{split}
\hat{x}_{k+1} &= A\hat{x}_k + Bu_k \\
&+ \rho P_{k+1|k+1} C^\top R^{-1} (y_{k+1} - C(A\hat{x}_k + Bu_k))
\end{split}
\end{align}
with $\hat{x}_0$ as the initial condition and the non-negative real number $\delta_k^2$ is given by the equation:
\begin{align}
\label{eq.sse_4}
\begin{split}
\delta_{k+1}^2 &= \, (1-\beta)(1-\rho)\delta_k^2 \\
&+ (y_{k+1} - C(A\hat{x}_k + Bu_k))[(1-\rho)^{-1}CP_{k+1|k}C^\top \\
&+ \rho^{-1}R]^{-1} (y_{k+1} - C(A\hat{x}_k + Bu_k)), \\
\delta_0^2 &= 0,
\end{split}
\end{align}
where $\beta$, $\rho$ are parameters with $0<\beta<1$ and $0<\rho<1$.

We point out two desirable properties of the state estimation algorithm given by~\cref{eq.sse_1,eq.sse_2,eq.sse_3,eq.sse_4}:
\begin{enumerate}[\itshape i)]
\item
The matrix $P_{k|k}$ does not depend on the outputs along the trajectory, and hence can be precomputed.
\item
The matrix $P_{k|k}$ converges to a steady state $P_\infty$, see~\cite{bertsekas1971recursive}.
\end{enumerate}
The first property will be helpful to calculate the bounds on uncertainty within the prediction horizon, and the second property will be beneficial to develop a time-invariant tightened constraint in steady state.

\subsubsection{State decomposition}
We decompose the state $x_k$ into two components instead of three components as in the prior work and as discussed in~(\ref{eq:state-decomposition-three}). The first component is the nominal state $\bar{x}_k$ as defined in~(\ref{eq:nominal-system}). The second component is the control error $s_k$, which is the difference between the real state $x_k$ and the nominal state $\bar{x}_k$:
\begin{equation}
\label{eq:s}
s_k = x_k - \bar{x}_k.
\end{equation}
The real state $x_k$ can then be represented by:
\begin{equation}
x_k = \bar{x}_k + s_k.
\end{equation}
We now tighten the constraint by considering the uncertainty in the error state $s_k$.

\subsubsection{Constraint tightening}
We use the same control input as in~(\ref{eq:control-input}). From~\cref{eq:system-x,eq:nominal-system,eq:control-input,eq:e,eq:s}, the dynamics for the control error $s_k$ is governed by:
\begin{equation}
s_{k+1} = (A + BK)s_k + w_k - BKe_k.
\end{equation}
Using the set-membership state estimation~(\ref{eq.sse_1}) yields the set recursion:
\begin{equation}
\label{eq:set-recursion-s-cl}
\mathbb{S}_{k+1} = (A + BK) \mathbb{S}_k \oplus \mathbb{W} \oplus -BK \mathbb{E}_{k|k},
\end{equation}
where
\begin{equation}
\label{eq:ellipsoids-e}
\mathbb{E}_{k|k} = \{e \mid e^\top P_{k|k}^{-1} e \le 1 - \delta_k^2\}.
\end{equation}
The sets $\mathbb{S}_k$ are bounds on the worst case evolution of $s_k$ starting from any $s_0$ within $\mathbb{S}_0 = \mathbb{E}_{0|0}$. Therefore, the constraint~(\ref{ineq:constraints}) can be robustly satisfied if the following inequality holds for any $s_k \in \mathbb{S}_k$, $e_k \in \mathbb{E}_{k|k}$, and $k \ge 0$:
\begin{equation}
\label{ineq:cstr-tightening-ellipsoids}
F\bar{x}_k + G\bar{u}_k + (F + GK)s_k - GKe_k \le f.
\end{equation}

The tightened constraint~(\ref{ineq:cstr-tightening-ellipsoids}) cannot be directly used in predictive control because $\mathbb{S}_k$ and $\mathbb{E}_{k|k}$ cannot be precomputed. In detail, $\delta_k^2$ in~(\ref{eq:ellipsoids-e}) depends on $y_k$, which is unknown before the time $k$. To overcome the problem, we present the following proposition bounding the estimation error $e_k$ without prior knowledge of the subsequent system outputs.
\begin{proposition}[Bounds on estimation error]{\ \\}
\label{prop.1}
Consider the set-membership state estimation~\cref{eq.sse_1,eq.sse_2,eq.sse_3,eq.sse_4} associated with the system~(\ref{eq:system}). At time $k$, it is guaranteed that the estimation error in the next $i$ steps $e_{k+i} \in \mathbb{E}_{k+i|k} \coloneqq \{e \mid e^\top P_{k+i|k+i}^{-1} e \le 1 - (1-\beta)^i(1-\rho)^i \delta_k^2\}$ for any $k \ge 0$, $i \ge 0$, and admissible disturbance sequences $\mathbf{w}$ and $\mathbf{v}$, where $\mathbb{E}_{k+i|k}$ is the bounding ellipsoid of the estimation error at time $k+i$ based on the information available at time $k$.
\end{proposition}
The proof is provided in Appendix~\ref{proof:prop-1}.

Note that the bounding set $\mathbb{E}_{k+i|k}$ of the estimation error $e_{k+i}$ can be precomputed at time $k$ for all $k \ge 0$ and $i \ge 0$, regardless of the actual subsequent disturbances and controls in the next $i$ steps.

To propagate the error within the prediction at time $k$, one can now bound $s_{k+i}$ by $\mathbb{S}_{k+i|k}$ with the dynamics of:
\begin{align}
\label{eq:set-recursion-s-pred}
\begin{split}
\mathbb{S}_{k+i+1|k} &= (A + BK) \mathbb{S}_{k+i|k} \oplus \mathbb{W} \oplus -BK \mathbb{E}_{k+i|k}, \\
\mathbb{S}_{k|k} &= \mathbb{S}_k,
\end{split}
\end{align}
where $\mathbb{S}_k$ is governed by dynamics~(\ref{eq:set-recursion-s-cl}). At time $k$, we can then derive the following tightened constraint within the prediction for any $s_{k+i} \in \mathbb{S}_{k+i|k}$, $e_{k+i} \in \mathbb{E}_{k+i|k}$, $i \ge 0$:
\begin{equation}
\label{ineq:cstr-tightening-ellipsoids-conservative}
F \bar{x}_{k+i} + G \bar{u}_{k+i} + (F + GK)s_{k+i} - GKe_{k+i} \le f.
\end{equation}

\begin{remark}[Constraint tightening for predictive control]
Any nominal state trajectory $\{\bar{x}_k, \, \bar{x}_{k+1}, \, \dots\}$ and nominal input sequence $\{\bar{u}_k, \, \bar{u}_{k+1}, \, \dots\}$ from time $k$ satisfying the constraint~(\ref{ineq:cstr-tightening-ellipsoids-conservative}) will also satisfy the constraint~(\ref{ineq:cstr-tightening-ellipsoids}) due to the fact that $\mathbb{E}_{k+i|k+i} \subseteq \mathbb{E}_{k+i|k}$ and $\mathbb{S}_{k+i} \subseteq \mathbb{S}_{k+i|k}$ for all $k \ge 0$ and $i \ge 0$ from Proposition~\ref{prop.1}. Therefore, we can use the constraint~(\ref{ineq:cstr-tightening-ellipsoids-conservative}), which can be precomputed, in predictive control to ensure the robust satisfaction of the original constraint~(\ref{ineq:constraints}).
\end{remark}

\begin{remark}[Computation of the tightened constraints]{\ \\}
There is no need to compute the sets $\mathbb{S}_k$ and $\mathbb{S}_{k+i|k}$ explicitly. Instead, one can compute the constraint tightening $(F + GK)s_{k+i} - GKe_{k+i}, \, \forall s_{k+i} \in \mathbb{S}_{k+i|k}, \, \forall e_{k+i} \in \mathbb{E}_{k+i|k}$ in~(\ref{ineq:cstr-tightening-ellipsoids-conservative}) by unrolling $\mathbb{S}_{k+i|k}$ and $\mathbb{S}_k$ over time and solving quadratically constrained linear programs.
\end{remark}

Alternatively, a time-invariant constraint tightening can be used after some time steps by adding some conservatism. In detail, if the state estimation enters steady state and $\mathbb{E}_{k|k}$ is over-approximated by $\mathbb{E}_{k|k} \subseteq \tilde{\mathbb{E}}_\infty \coloneqq \{e \mid e^\top P_\infty^{-1} e \le 1\}$, the dynamics~(\ref{eq:set-recursion-s-cl}) and~(\ref{eq:set-recursion-s-pred}) become identical as:
\begin{equation}
\label{eq:set-recursion-s-ss}
\mathbb{S}_{k+1} = (A + BK) \mathbb{S}_k \oplus \mathbb{W} \oplus -BK \tilde{\mathbb{E}}_\infty.
\end{equation}
For any initial $C$ set $\mathbb{S}_0$, the sets $\mathbb{S}_k$ governed by~(\ref{eq:set-recursion-s-ss}) converge in the Hausdorff metric to the minimal RPI set $\mathbb{S}_\infty$ satisfying:
\begin{equation}
\label{eq:s-mRPI}
\mathbb{S}_\infty = (A + BK) \mathbb{S}_\infty \oplus \mathbb{W} \oplus -BK \tilde{\mathbb{E}}_\infty,
\end{equation}
which allows to bound $\mathbb{S}_k$ by computing only a finite number of sets as in polytope settings~\cite{mayne2006robust,mayne2009robust,kogel2017robust,lorenzetti2020simple,kogel2021robust}.

Under \textit{ellipsoidal} uncertainty $\mathbb{W}$ and $\tilde{\mathbb{E}}_\infty$, instead of expressing the minimal RPI set $\mathbb{S}_\infty$ as a \textit{polytope}, we propose to directly calculate the time-invariant tightened constraint using quadratically constrained quadratic program without explicit computation of $\mathbb{S}_\infty$. The proposed algorithm avoids the usage of Minkowski sum, thus scaling well to high-dimensional systems. The details are provided in Appendix~\ref{appendix.B}.
\section{Closed Loop Properties}
In this section, we present conditions on the MPC setup to guarantee robust constraint satisfaction and robust stability based on the tightened constraint~(\ref{ineq:cstr-tightening-ellipsoids-conservative}) from Section~\ref{sec:cstr-tightening}.

In the following we present the optimization problem underlying the MPC. To employ the constraint tightening with reduced conservatism from Section~\ref{sec:cstr-tightening}, the proposed robust output feedback approach uses a similar controller structure to the one in~\cite{kogel2017robust,lorenzetti2020simple}, which is in contrast to~\cite{mayne2006robust,mayne2009robust,kogel2021robust}: The nominal state $\bar{x}_k$ for $k>0$ is not determined by the optimization problem, instead it is given by the dynamics~(\ref{eq:nominal-system}) with the previous nominal state/input $\bar{x}_{k-1}/\bar{u}_{k-1}$.

The optimal control problem at time $k$ is given by:
\begin{align}
\label{prob:Safe-SM-MPC}
\begin{split}
&V_k^* = \min_{\bar{\mathbf{x}}_k, \bar{\mathbf{u}}_k} \sum_{i=0}^{N-1} q(\bar{x}_{k+i|k}, \bar{u}_{k+i|k}) + p(\bar{x}_{k+N|k}) \\
&\text{s.t.} \; \; \bar{x}_{k+i+1|k} = A\bar{x}_{k+i|k} + B\bar{u}_{k+i|k} \\
&\quad \; \; F\bar{x}_{k+i|k} + G\bar{u}_{k+i|k} \le f - (F+GK)s_{k+i} + GKe_{k+i}, \\
&\qquad \qquad \qquad \qquad \qquad \; \; \forall s_{k+i} \in \mathbb{S}_{k+i|k}, \; \forall e_{k+i} \in \mathbb{E}_{k+i|k}, \\
&\quad \; \; \bar{x}_{k+N|k} \in \mathbb{X}_k^f, \; \bar{x}_{k|k} = \bar{x}_k, \\
\end{split}
\end{align}
where $i=0, \dots N-1$, $\bar{\mathbf{x}}_k = \{ \bar{x}_{k|k}, \dots \bar{x}_{k+N|k} \}$ is the planned nominal state trajectory, $\bar{\mathbf{u}}_k = \{ \bar{u}_{k|k}, \dots \bar{u}_{k+N-1|k} \}$ is the planned input sequence, $N$ is the horizon, $q(\cdot,\cdot)$ is a positive definite stage cost, $p(\cdot)$ is a positive definite terminal cost, $\mathbb{S}_{k+i|k}$ is the bounding set of the control error governed by~(\ref{eq:set-recursion-s-pred}), $\mathbb{E}_{k+i|k}$ is the bounding set of the estimation error governed by Proposition~\ref{prop.1}, $\mathbb{X}_k^f$ is the terminal set, and $\bar{x}_k$ is governed by dynamics~(\ref{eq:nominal-system}) with the initial condition of $\bar{x}_0=\hat{x}_0$ to achieve less conservative constraint tightening as described in Section~\ref{sec:cstr-tightening}.

Upon solving~(\ref{prob:Safe-SM-MPC}) at time $k$, the controller applies
\begin{equation}
\label{eq:Safe-SM-MPC-controller}
u_k = \bar{u}_{k|k}^* + K(\hat{x}_k - \bar{x}_k)
\end{equation}
to the system~(\ref{eq:system}), where $\hat{x}_k$ is the state estimate given by~(\ref{eq.sse_3}), and $\bar{u}_{k|k}^*$ is the first part of the optimal nominal input from~(\ref{prob:Safe-SM-MPC}). The controller~(\ref{eq:Safe-SM-MPC-controller}) along with the state estimation~(\ref{eq.sse_3}) forms the receding horizon control strategy for the system.

For the brevity of notation, let $\bar{f}_k$ be the tightened constraint at the end of the horizon in~(\ref{prob:Safe-SM-MPC}):
\begin{align}
\begin{split}
\bar{f}_k = &f - \max_{s_{k+N}} (F+GK)s_{k+N} - \max_{e_{k+N}} - GKe_{k+N}. \\
\text{s.t.} \quad &s_{k+N} \in \mathbb{S}_{k+N|k}, \; e_{k+N} \in \mathbb{E}_{k+N|k},
\end{split}
\end{align}
where the maximization is performed elementwise, compare~\cite{Kouvaritakis2015model}. We introduce the following assumption on the terminal cost and terminal sets, compare~\cite{mayne2009robust,kogel2017robust}:
\begin{assumption}[Terminal cost and sets]{\ \\}
\label{assumption-1}
There exist a terminal cost $p(\cdot)$ and terminal sets $\mathbb{X}_k^f$ such that:
\begin{subequations}
\begin{align}
p\left((A + BK)x\right) - p(x) \le -q(x, Kx), \; \forall x \in \mathbb{X}_k^f, \\
(A + BK) x \in \mathbb{X}_{k+1}^f, \; (F+GK)x \le \bar{f}_k, \; \forall x \in \mathbb{X}_k^f.
\end{align}
\end{subequations}
\end{assumption}

We now establish the recursive feasibility and stability of the proposed robust, output feedback MPC:
\begin{theorem}[Closed loop properties]{\ \\}
\label{theorem-1}
Let Assumption~\ref{assumption-1} hold. If the problem~(\ref{prob:Safe-SM-MPC}) is feasible for time $k=0$, then the closed-loop system given by~\cref{eq:system,eq:disturbances,eq.sse_3,prob:Safe-SM-MPC,eq:Safe-SM-MPC-controller} has the following properties for any admissible realization of the initial uncertainty $e_0$, the state disturbance sequence $\mathbf{w}$, and the output disturbance sequence $\mathbf{v}$:
\begin{itemize}
\item
Recursive feasibility:~(\ref{prob:Safe-SM-MPC}) is feasible for any $k > 0$,
\item
Constraint satisfaction: $Fx_k + Gu_k \le f$ for any $k \ge 0$,
\item
Convergence: the nominal state $\bar{x}_k$ is exponentially stable and $x_k$ converges to the set $\mathbb{S}_\infty$.
\end{itemize}
\end{theorem}
The proof is provided in Appendix~\ref{proof:theorem-1}.
\section{Simulation Examples}
We illustrate the results on two example systems: a double integrator with state dimension $n=2$ as in~\cite{mayne2006robust,mayne2009robust,kogel2021robust}, which shows the improved constraint tightening with reduced conservatism relative to prior works, and a quadrotor system with $n=12$, which shows that the proposed robust output feedback MPC can generalize to high-dimensional systems. Hereafter, we define the stage cost and terminal cost as:
\begin{equation*}
q(x,u) \coloneqq (1/2)[x^\top \tilde{Q} x + u^\top \tilde{R} u], \quad p(x) \coloneqq (1/2)x^\top \tilde{P} x,
\end{equation*}
where $\tilde{Q}$, $\tilde{R}$, and $\tilde{P}$ are positive definite matrices.

\subsection{Double Integrator}
To illustrate the improved constraint tightening we consider the following system~\cite{mayne2006robust,mayne2009robust}:
\begin{equation*}
x_{k+1}
=
\begin{bmatrix}
1 & 1 \\
0 & 1
\end{bmatrix} x_k
+
\begin{bmatrix}
1 \\
1
\end{bmatrix} u_k
+ w_k, \quad y_k
=
\begin{bmatrix}
1 & 1
\end{bmatrix}x_k + v_k,
\end{equation*}
where the \textit{ellipsoidal} disturbance bounds are given by:
\begin{align*}
\|w_k\|_2 \le \lambda, \, \lambda > 0, \quad \|v_k\|_2 \le \mu, \, \mu > 0.
\end{align*}
The state and input constraints are:
\begin{equation*}
x_k \in [-50, 3] \times [-50, 3], \quad u_k \in [-3, 3].
\end{equation*}

The control gain is $K=(-0.6136, -0.9962)$. The state estimation parameters $\beta$ and $\rho$ are determined offline by a grid search such that the trace of the matrix $P_\infty$ is minimized. For the baselines using polytopic sets for constraint tightening, we approximated the disturbance sets $\mathbb{W}$ and $\mathbb{V}$ by axis-aligned minimum bounding boxes and chose the observer gain $L=(-1, -1)$ as in~\cite{mayne2006robust}. For simplicity it is assumed that $\mathbb{E}_0 = \mathbb{E}_\infty$ and $\Xi_0=\{0\}$ is used for the polytope approaches and that $P_{0|0} = P_\infty$ is used for the proposed method.

\textit{Constraint tightening:}
\begin{figure}[t]
  \centering
  \begin{subfigure}[b]{0.78\linewidth}
    \includegraphics[width=\linewidth]{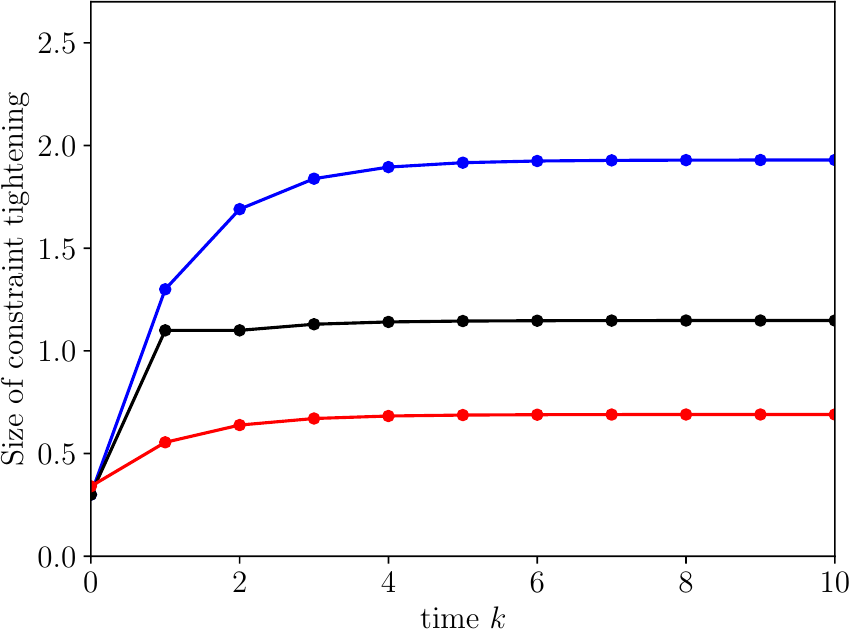}
    \caption{Constraint tightening for $x[1]$.}
    \vspace{2mm}
  \end{subfigure}
  \begin{subfigure}[b]{0.78\linewidth}
    \includegraphics[width=\linewidth]{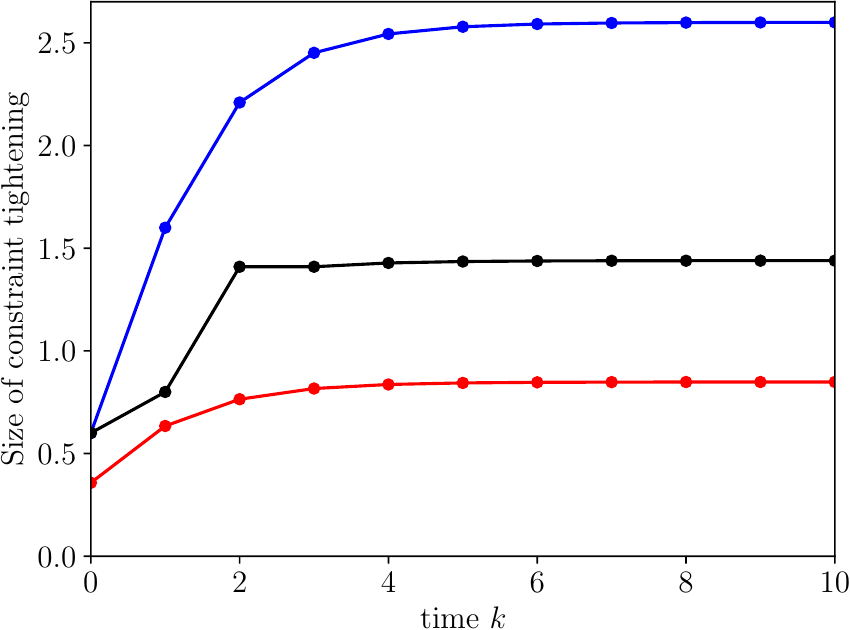}
    \caption{Constraint tightening for $x[2]$.}
    \vspace{2mm}
  \end{subfigure}
  \begin{subfigure}[b]{0.78\linewidth}
    \includegraphics[width=\linewidth]{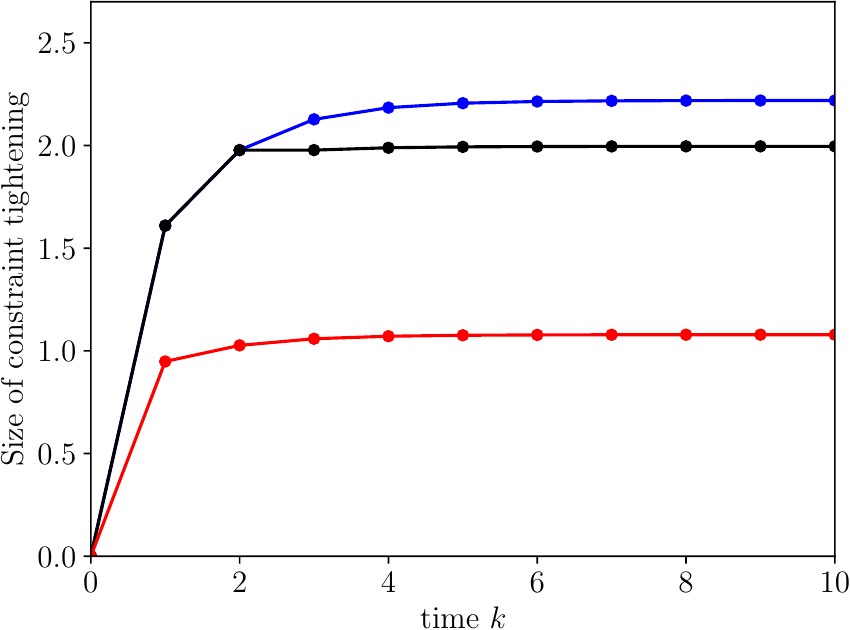}
    \caption{Constraint tightening for $u$.}
  \end{subfigure}
  \caption{Constraint tightening using the two set approach~(\ref{ineq:cstr-tightening-separate-sets}) (blue), the single set approach~(\ref{ineq:cstr-tightening-single-set}) (black), and the proposed approach~(\ref{ineq:cstr-tightening-ellipsoids-conservative}) (red).}
  \label{fig:cstr-tightening}
  \vspace{-3mm}
\end{figure}
We set $\lambda=0.1$ and $\mu=0.05$. Figure~\ref{fig:cstr-tightening} illustrates the constraint tightening on $x[1]$, $x[2]$, and $u$ computed at time $k=0$ using the proposed approach~\cref{eq:set-recursion-s-pred,ineq:cstr-tightening-ellipsoids-conservative}, the two set approach~\cref{eq:separate-sets-recursion-E,eq:separate-sets-recursion-Xi,ineq:cstr-tightening-separate-sets}, and the single set approach~\cref{eq:single-set-recursion-Z,ineq:cstr-tightening-single-set}. We observe that the proposed approach delivers significantly less conservative constraint tightening by avoiding the over-approximation of ellipsoidal uncertainty. The two set approach and the single set approach produce an identical constraint tightening for time $k=0$ due to the same initialization of $\mathbb{E}_k$ and $\Xi_k$.

We further investigate the constraint tightening in steady state with $\mathbb{E}_\infty$ and $\Xi_\infty$ in the two set approach, $\mathbb{Z}_\infty$ in the single set approach, and $\mathbb{S}_\infty$ and $\tilde{\mathbb{E}}_\infty$ in the proposed approach. Disturbance bounds with $\lambda=\mu=0.25$ are used. Table~\ref{table:cstr-tighteing-ti} summarizes the results on respective states and control input. The constraint tightening will converge to the values listed in the table as $k \to \infty$. Note that the two set approach and the single set approach generate a constraint tightening on $u$ of $3.884$ and $3.447$ respectively, which both exceed the radius of the interval of the constraint on $u$, thus making the robust control infeasible. In contrast, the proposed approach generates a much smaller constraint tightening on $u$ of $1.963$ due to the reduced conservatism. In such cases, only the proposed controller can robustly stabilize the system while ensuring the constraint satisfaction.
\begin{table}[t]
  \begin{center}
    \caption{Time invariant constraint tightening}
    \label{table:cstr-tighteing-ti}
    \begin{tabular}{ l | c  c  c }
      \toprule
      Approach & $x[1]$ & $x[2]$ & $u$ \\
      \midrule
      \rule{-2.5pt}{2ex} Two set~\cref{eq:separate-sets-recursion-E,eq:separate-sets-recursion-Xi,ineq:cstr-tightening-separate-sets} & $3.352$ & $4.500$ & $3.884$ \\
      Single set~\cref{eq:single-set-recursion-Z,ineq:cstr-tightening-single-set} & $1.712$ & $2.294$ & $3.447$ \\
      Proposed~\cref{eq:set-recursion-s-pred,ineq:cstr-tightening-ellipsoids-conservative} & $1.174$ & $1.443$ & $1.963$ \\
      \bottomrule
    \end{tabular}
  \end{center}
  \vspace{-4mm}
\end{table}

\textit{Simulation:} The closed loop performance of the proposed approach and the single set approach of K{\"o}gel et al~\cite{kogel2017robust} is illustrated using simulations.

For the cost function we choose $N=15$, $\tilde{Q}=I$, $\tilde{R}=0.01$, and $\tilde{P}$ to be the solution to the algebraic Riccati equation of the infinite horizon LQR. The terminal set is computed as a maximal positive invariant set for the nominal system. The disturbances are randomly sampled from the bounding sets with $\lambda=0.1$ and $\mu=0.05$ and are kept identical for two approaches.
\begin{figure}[t]
  \centering
  \begin{subfigure}[b]{0.46\linewidth}
    \includegraphics[width=\linewidth]{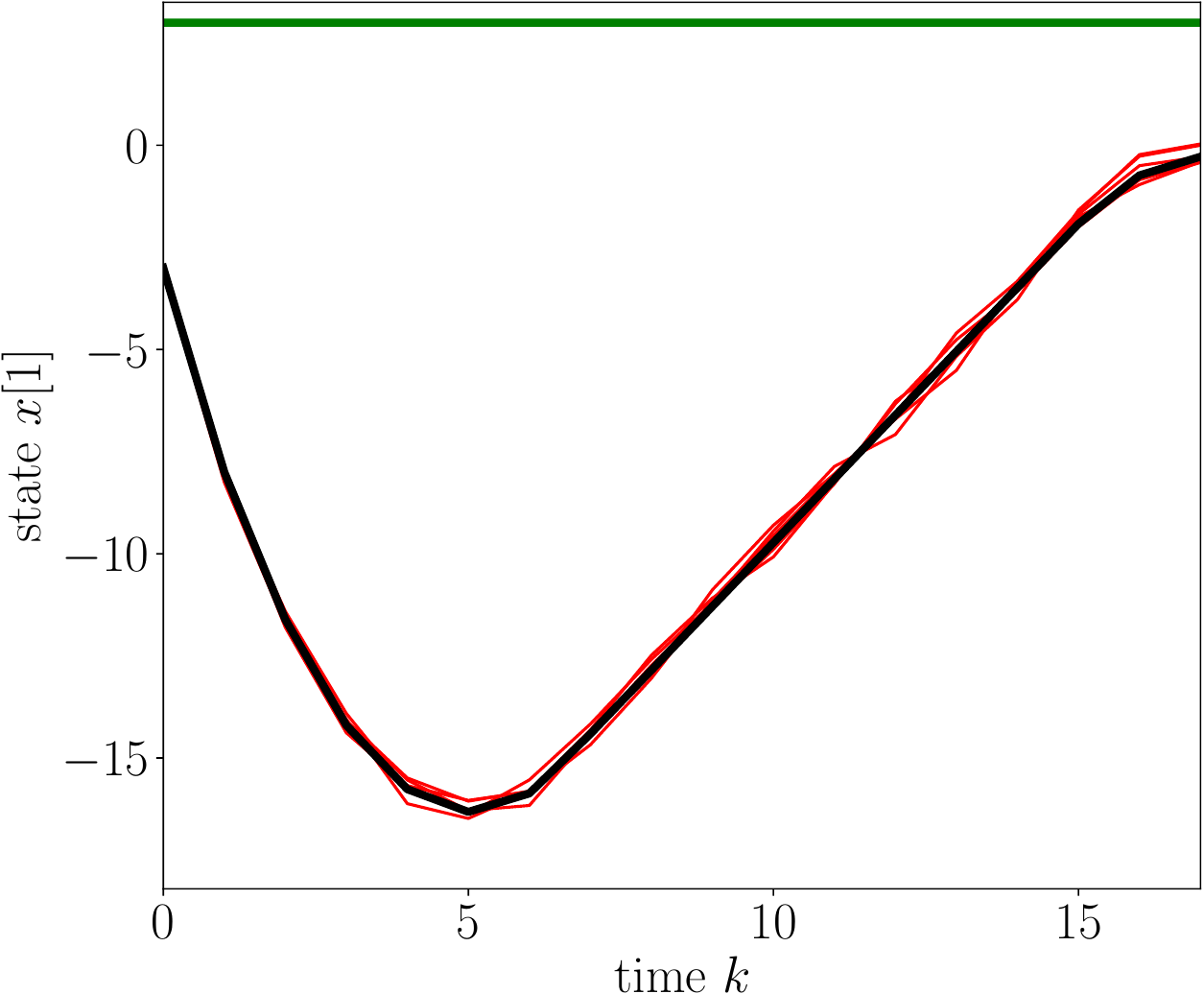}
  \end{subfigure}
  \begin{subfigure}[b]{0.46\linewidth}
    \includegraphics[width=\linewidth]{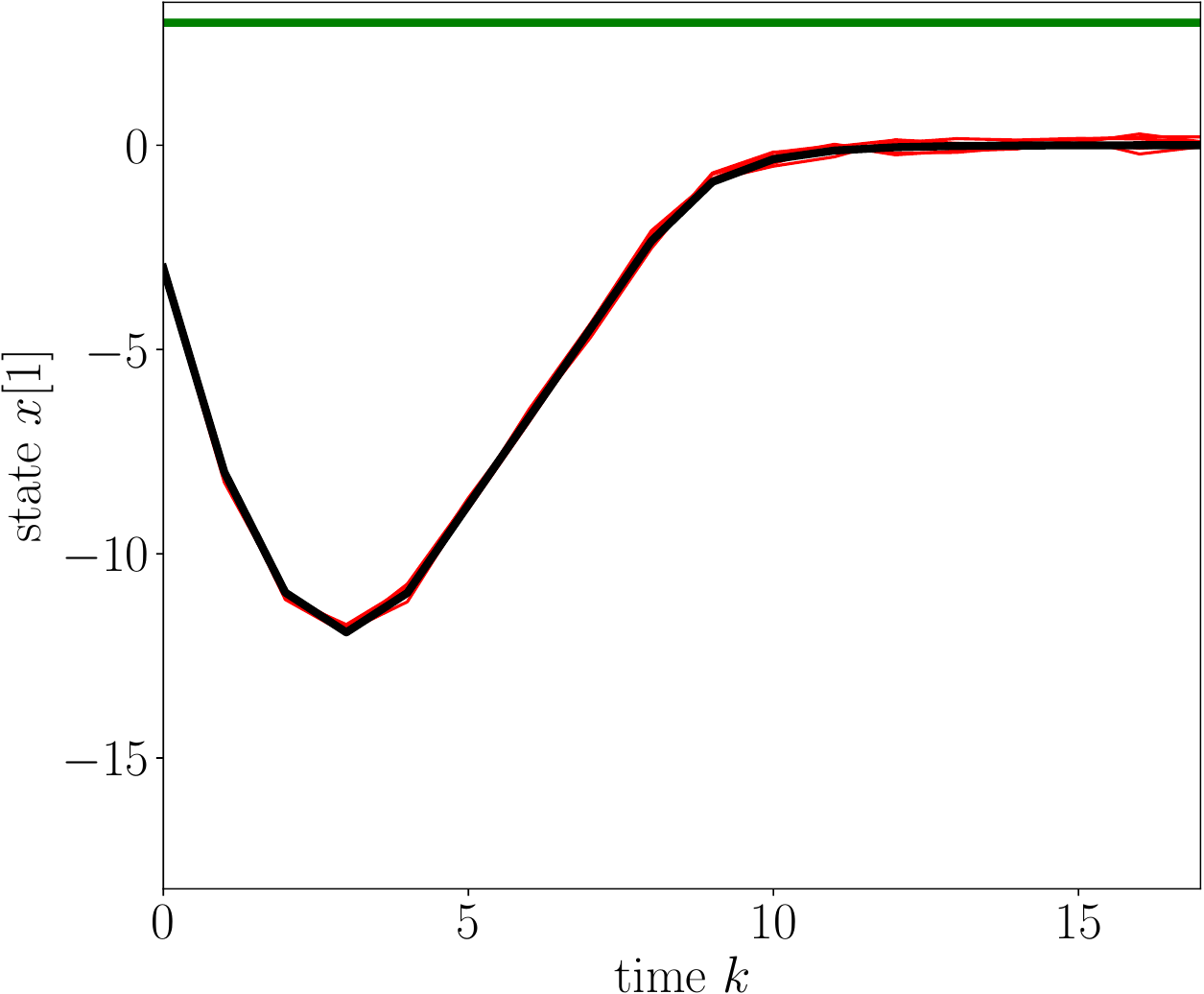}
  \end{subfigure}
  \begin{subfigure}[b]{0.46\linewidth}
    \includegraphics[width=\linewidth]{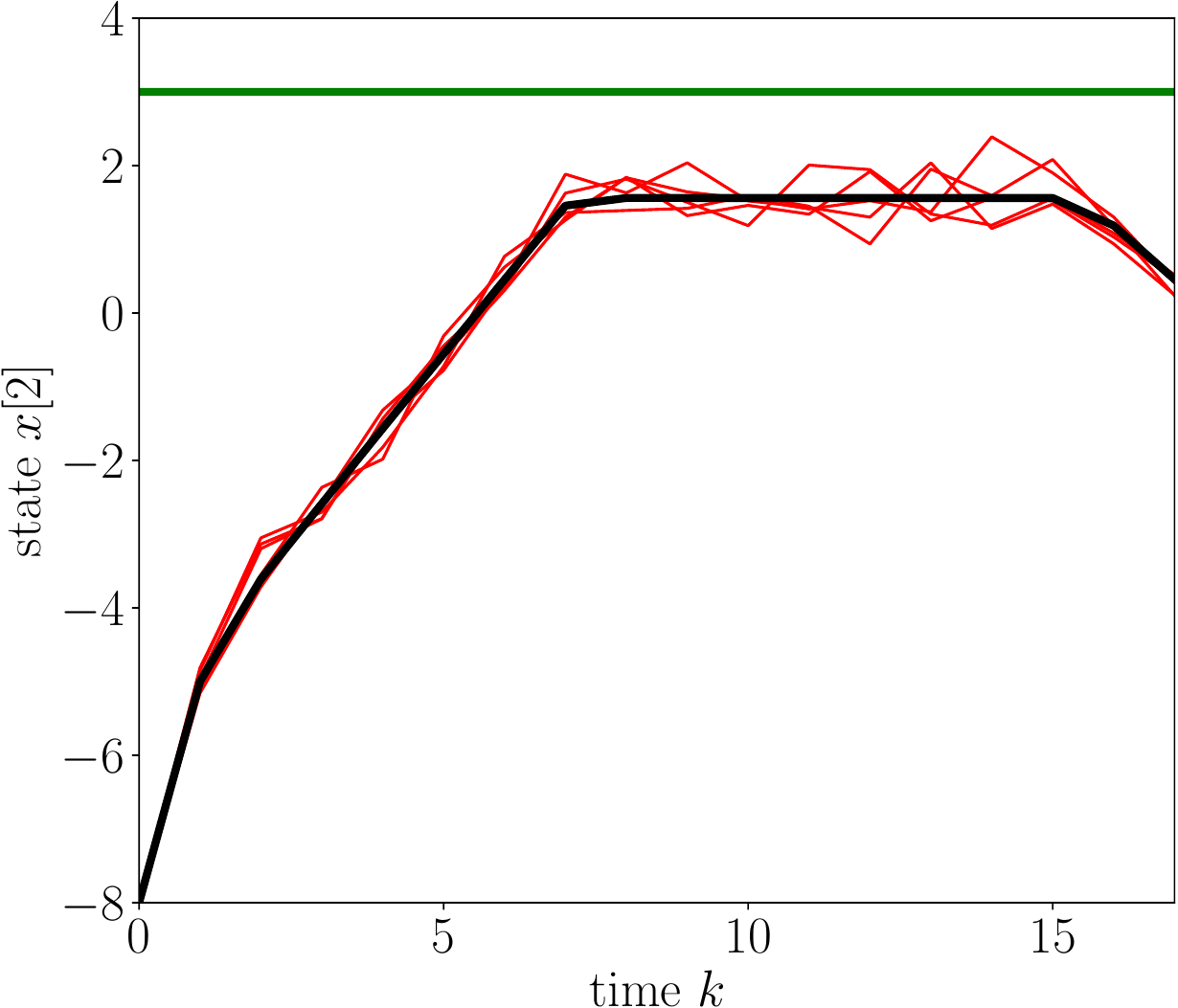}
  \end{subfigure}
  \begin{subfigure}[b]{0.46\linewidth}
    \includegraphics[width=\linewidth]{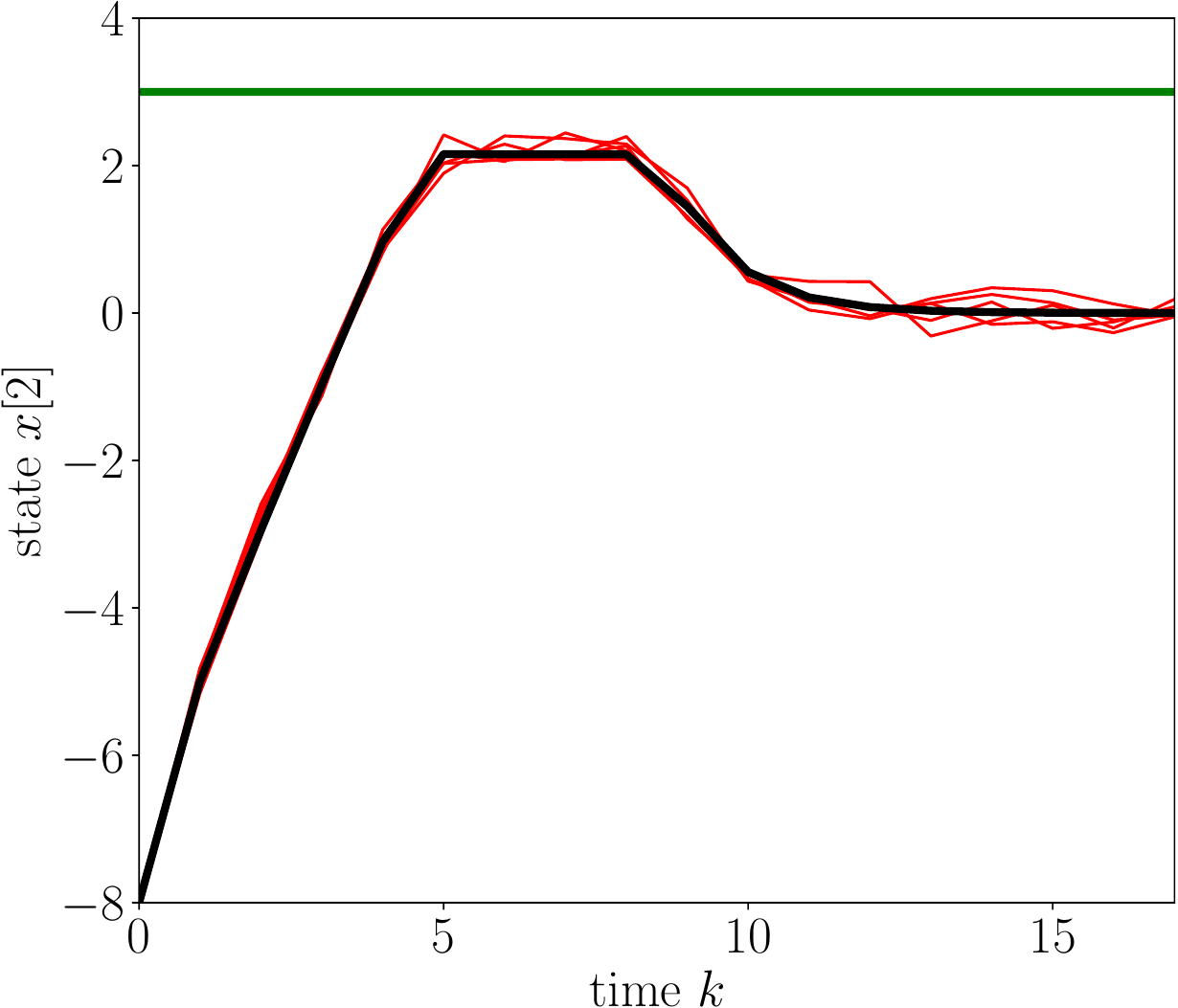}
  \end{subfigure}
  \begin{subfigure}[b]{0.46\linewidth}
    \includegraphics[width=\linewidth]{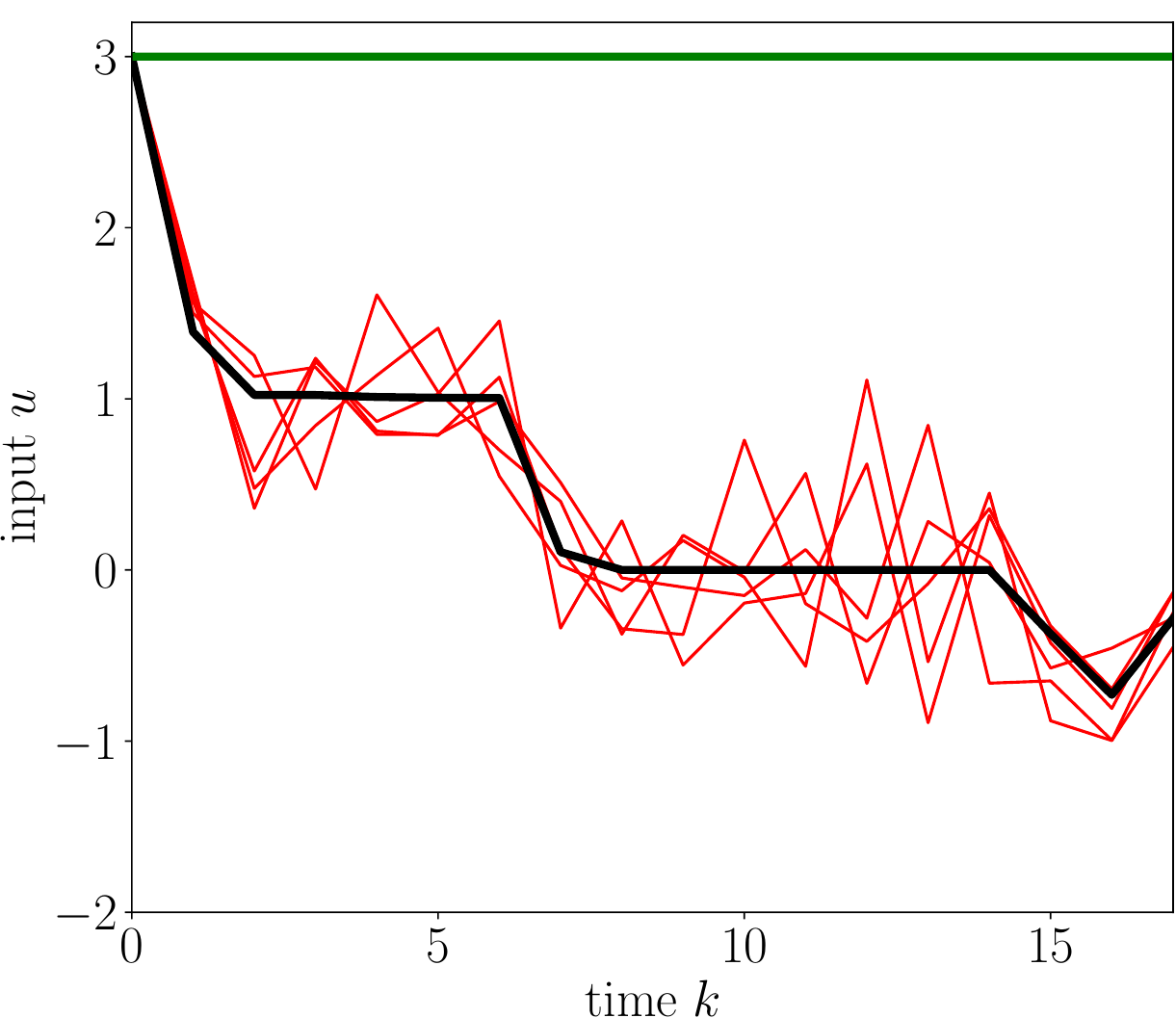}
  \end{subfigure}
  \begin{subfigure}[b]{0.46\linewidth}
    \includegraphics[width=\linewidth]{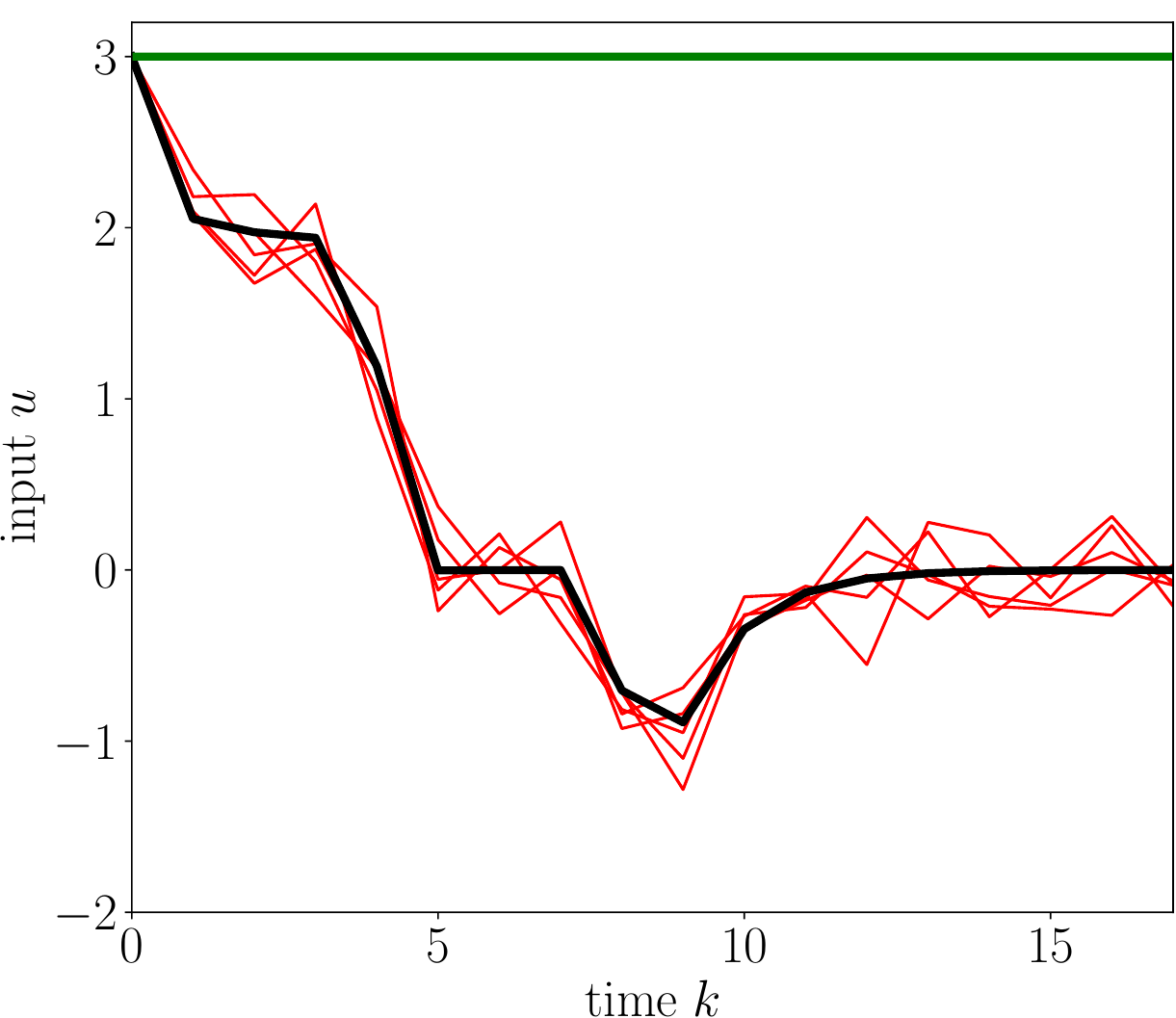}
  \end{subfigure}
  \caption{Closed-loop response of the double integrator. \textit{Left}: single set approach based on K{\"o}gel et al~\cite{kogel2017robust}. \textit{Right}: proposed approach. \textit{Black}: nominal path with $w_k=0$, $v_k=0$. \textit{Red}: sample paths for random noise with uniform distribution. \textit{Green}: Constraints on the real state and input.}
  \label{fig:double-integrator}
  \vspace{-3mm}
\end{figure}

Figure~\ref{fig:double-integrator} shows the closed-loop response of both approaches under uncertainty, starting from the initial state $x_0 = (-3.1, -8)^\top$ and the initial state estimate $\hat{x}_0 = (-3, -8)^\top$. Note that the state and input constraints are satisfied for both approaches. Moreover, the proposed approach allows to move the state $x[2]$ closer to the boundary of the constraint than the single set approach. A larger input $u$ is also allowed in the first few time steps due to the reduced conservatism using the proposed approach, compare Figure~\ref{fig:cstr-tightening}. As a result, a faster convergence to the origin is achieved compared to the single set approach.

\subsection{Quadrotor Dynamics}
To illustrate that the proposed method can scale up to high-dimensional systems we consider quadrotor dynamics with $12$ states and $4$ inputs~\cite{liu2020bio}:
\begin{align}
\label{eq.quadrotor-dynamics}
\begin{split}
\mathbf{\dot{p}} &= \mathbf{v}, \qquad \quad m\mathbf{\dot{v}} = mg\mathbf{e}_3 - T\mathbf{R}\mathbf{e}_3, \\
\mathbf{\dot{R}} &= \mathbf{R} \boldsymbol{\hat{\omega}}, \qquad J\boldsymbol{\dot{\omega}} = \mathbf{M} - \boldsymbol{\omega} \times J\boldsymbol{\omega},
\end{split}
\end{align}
where $\mathbf{p}=(p_x,p_y,p_z)^\top$ is the position, $\mathbf{R} \in SO(3)$ is rotation matrix representing the quadrotor attitude corresponding to the 3-2-1 Euler angle $\mathbf{\Omega}=(\phi, \theta, \psi)^\top$, $\mathbf{v}=(v_x,v_y,v_z)^\top$ is the translational velocity, $\boldsymbol{\omega}=(\omega_\phi, \omega_\theta, \omega_\psi)^\top$ is the angular velocity, $T$ is the total thrust, $\mathbf{M}=(M_x, M_y, M_z)^\top$ is the moment, $\hat{\cdot}: \mathbb{R}^3 \to SO(3)$ is the hat operator, $m$ is the mass of the quadrotor, $g$ is the gravitational force, and $J = \text{diag}(J_x, \, J_y, \, J_z)$ is the moment of inertial matrix. The inertial property of the quadrotor model is adopted from~\cite[Chapter~16]{valavanis2015handbook}. The model used for control is the linearized model of~(\ref{eq.quadrotor-dynamics}) around the equilibrium state where $\mathbf{\Omega}_e=[0,0,0]^\top$ and $T_e=mg$. To apply MPC controllers, a time discretization is used with $dt=0.2 \,$s. The output matrix $C$ is defined as an identity matrix. State and output disturbances are added to the model to introduce the uncertainty.

The ellipsoidal disturbance bounds are given by:
\begin{align*}
\|w_k\|_2 \le 0.03, \quad \|v_k\|_2 \le 0.03.
\end{align*}
The state and input constraints are $\mathbf{\Omega}_k \in [-\pi/9, \pi/9] \times [-\pi/9, \pi/9] \times [-\pi/9, \pi/9]$ and $T_k \in [-5, 5]$.
\begin{figure}[t]
  \centering
  \begin{subfigure}[b]{0.45\linewidth}
    \includegraphics[width=\linewidth]{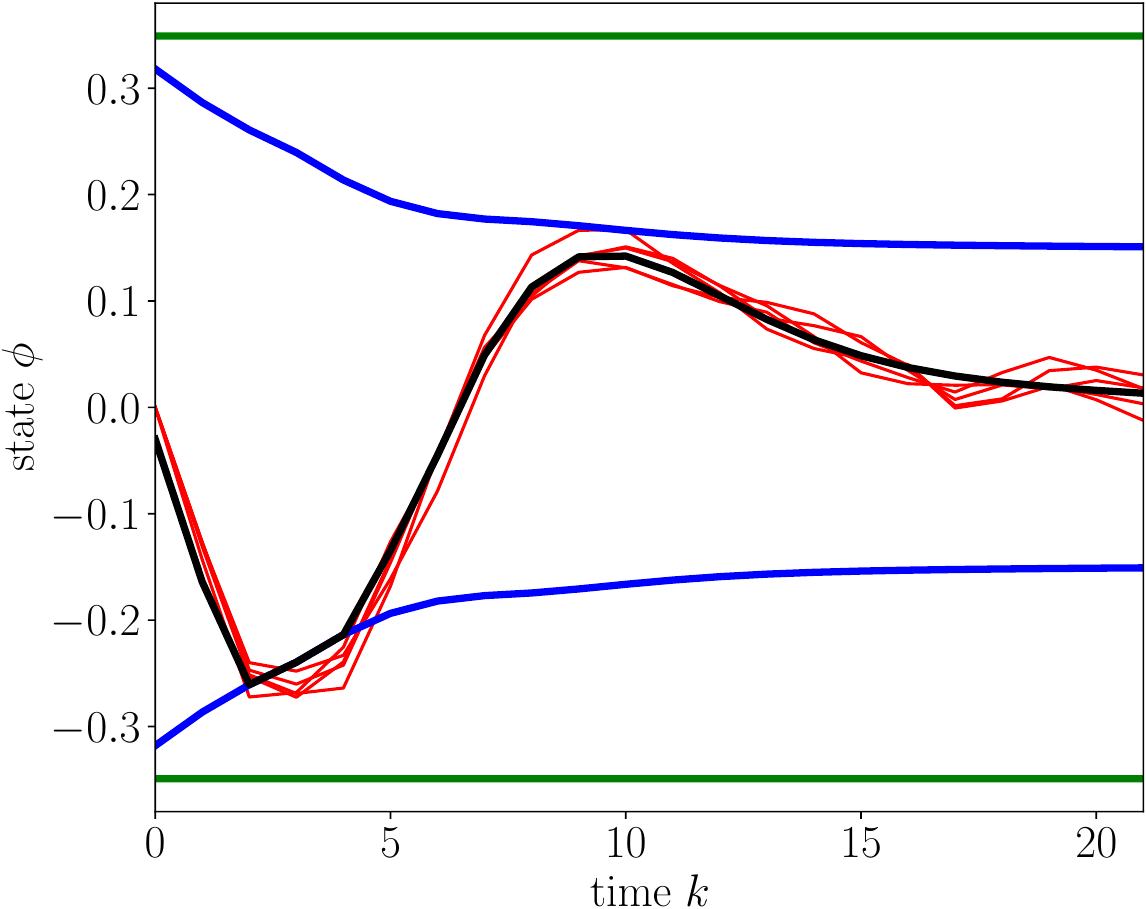}
  \end{subfigure}
  \begin{subfigure}[b]{0.45\linewidth}
    \includegraphics[width=\linewidth]{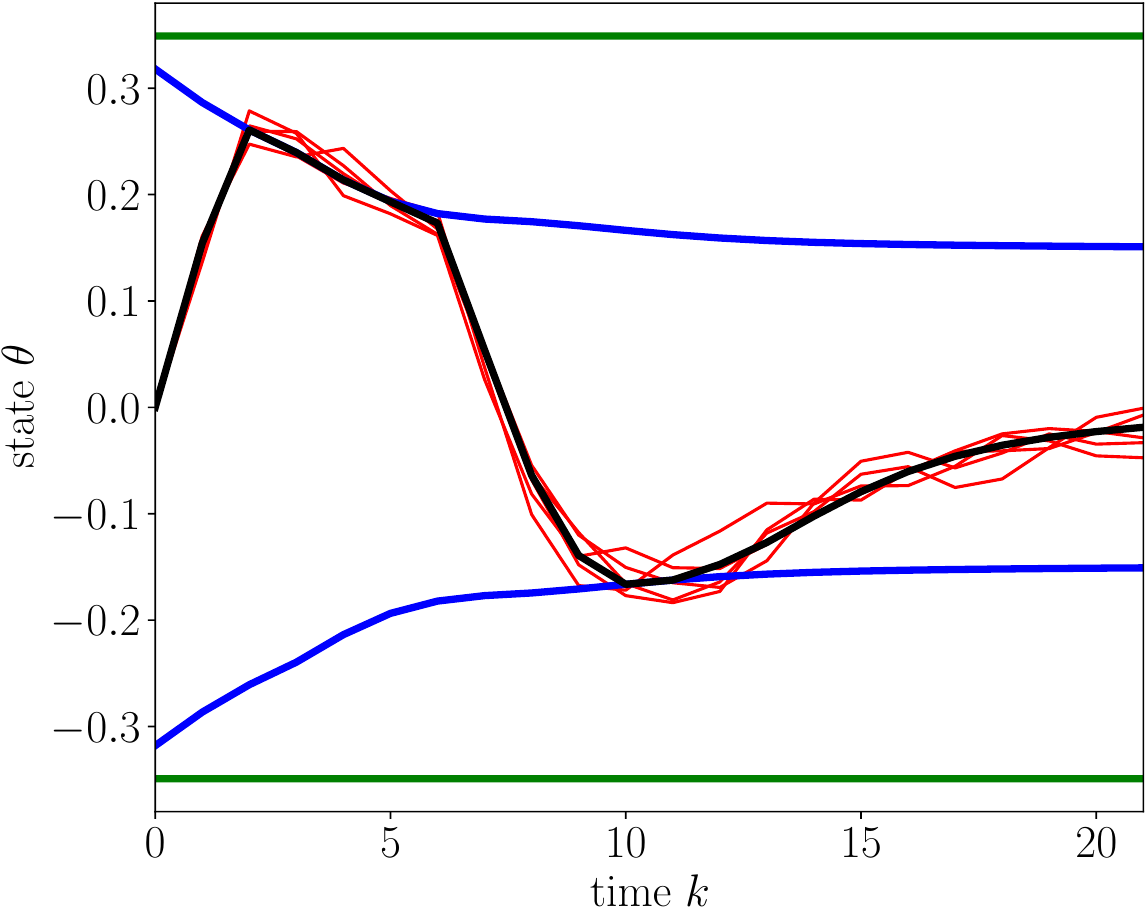}
  \end{subfigure}
  \begin{subfigure}[b]{0.44\linewidth}
    \includegraphics[width=\linewidth]{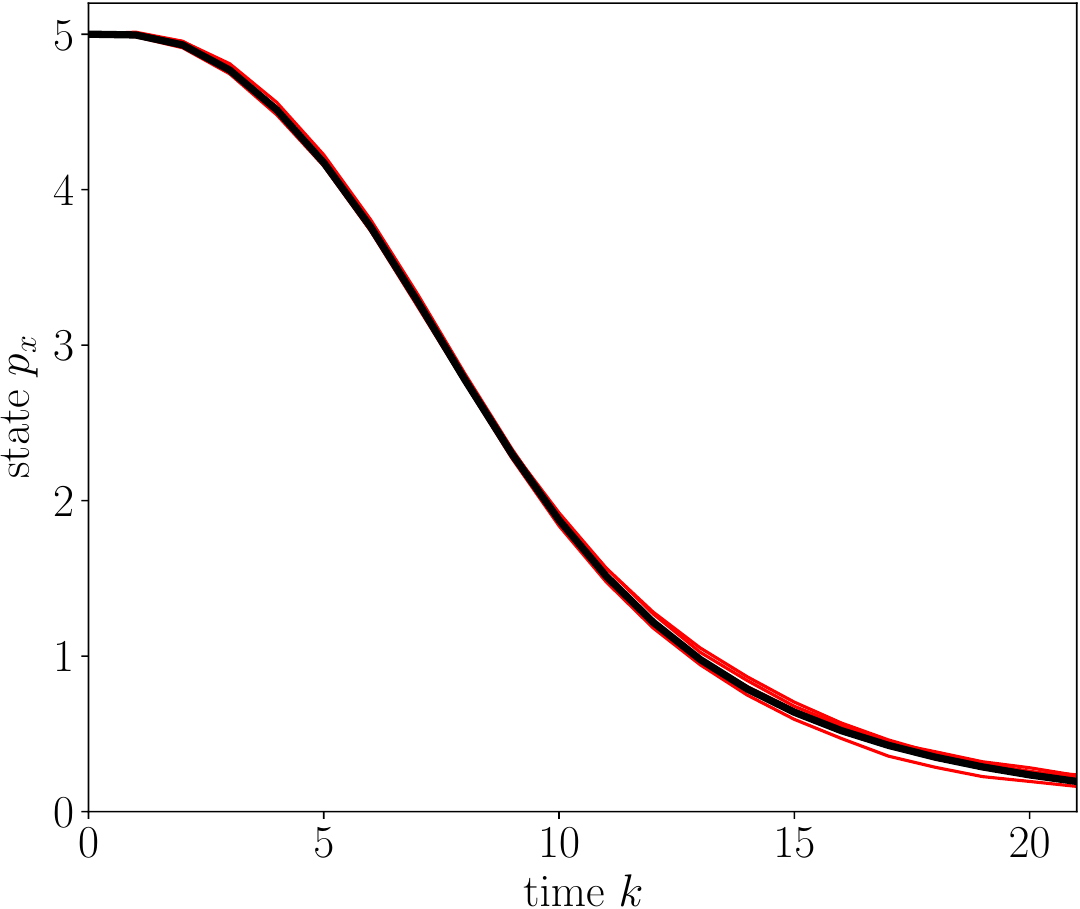}
  \end{subfigure} \hspace{1mm}
  \begin{subfigure}[b]{0.44\linewidth}
    \includegraphics[width=\linewidth]{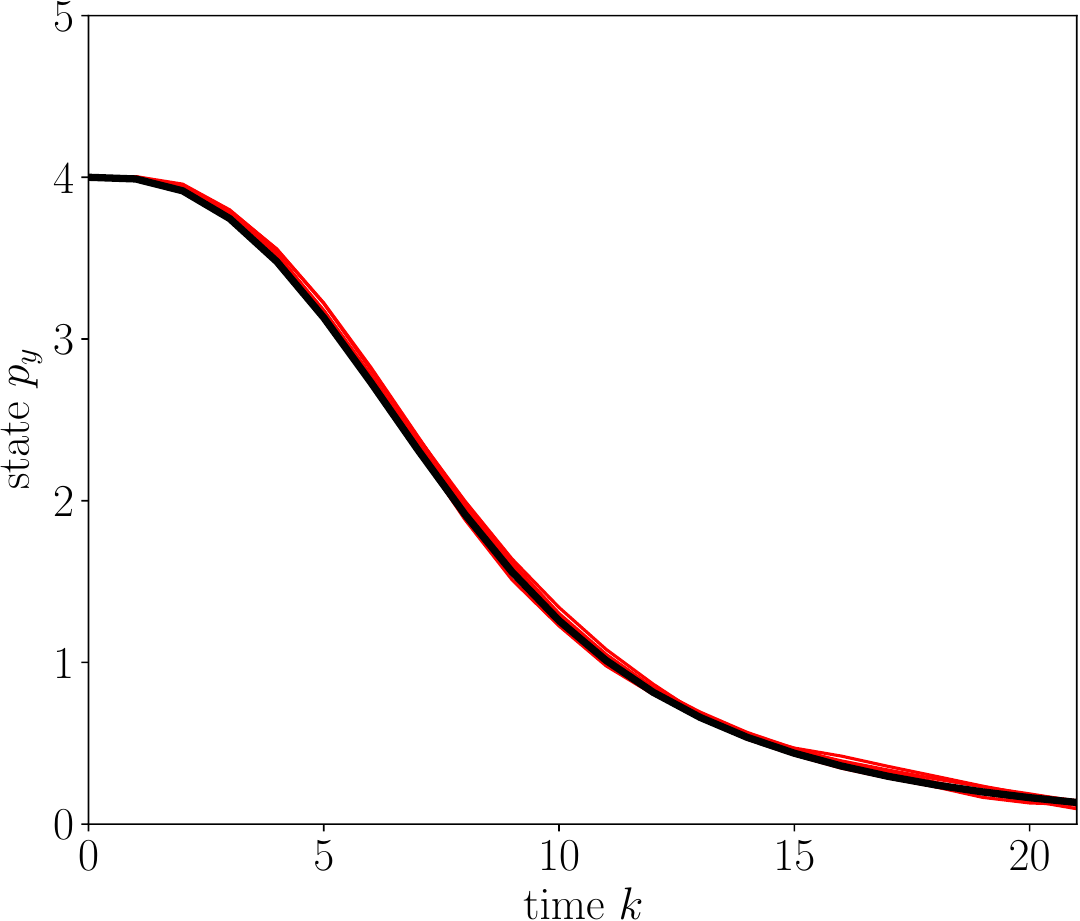}
  \end{subfigure}
  \caption{Closed-loop response of the quadrotor system using the proposed approach. \textit{Black}: nominal path with $w_k=0$, $v_k=0$. \textit{Red}: sample paths for random noise with uniform distribution. \textit{Green}: Constraints on the real state. \textit{Blue}: Tightened constraints on the nominal state.}
  \label{fig:quadrotor}
  \vspace{-3mm}
\end{figure}

Figure~\ref{fig:quadrotor} shows the closed-loop response of the proposed approach for a regulation task on the linearized model, starting from the initial state $\mathbf{p}_0=(5,4,0)^\top$, $\mathbf{\Omega}_0=(0,0,0)^\top$, $\mathbf{v}_0=(0,0,0)^\top$, $\boldsymbol{\omega}_0=(0,0,0)^\top$ and the initial state estimate $\hat{\mathbf{p}}_0=(5,4,0)^\top$, $\hat{\mathbf{\Omega}}_0=(-0.03,0,0)^\top$, $\hat{\mathbf{v}}_0=(0,0,0)^\top$, $\hat{\boldsymbol{\omega}}_0=(0,0,0)^\top$. We observe that the constraints are robustly satisfied, the MPC optimization problem~(\ref{prob:Safe-SM-MPC}) remains feasible, and the system is robustly stabilized. In contrast, the single set approach can be computationally challenging in high-dimensional systems due to explicit computation of mRPI sets, see~\cite[Remark~5]{kogel2017robust}.
\section{Conclusion}
We proposed a novel robust, output feedback model predictive controller for constrained linear systems with \textit{ellipsoidal} state and output disturbances. The approach combines a state estimator with a predictive control law and derives the constraint tightening directly from ellipsoidal sets. This avoids the conservatism introduced by the over-approximation of the ellipsoidal uncertainty by polytopes in previous works. Moreover, the proposed method does not require the explicit computation of minimal RPI sets and thus can scale up to high-dimensional systems efficiently. Conditions guaranteeing robust stability were outlined. Examples illustrated the advantages of the proposed approach. Future works include real-world experiments on LTI systems and an extension of the proposed method to nonlinear systems.

\bibliographystyle{IEEEtran}
\bibliography{BibFile}

\appendix

\subsection{Proof of Proposition~\ref{prop.1}}
\label{proof:prop-1}
\begin{proof}
Given the information available at time $k$, we first prove by induction that $\delta_{k+i}^2 \ge (1-\beta)^i(1-\rho)^i \delta_k^2$ for any $i \ge 0$. Clearly, $i=0$ satisfies the inequality. Assume that $\delta_{k+i}^2 \ge (1-\beta)^i(1-\rho)^i \delta_k^2$. From~(\ref{eq.sse_4}), we then have:
\begin{align*}
\delta_{k+i+1}^2 &= (1-\beta)(1-\rho)\delta_{k+i}^2 + (y_{k+i+1} \\
&\; \; - C(A\hat{x}_{k+i} + Bu_{k+i}))[(1-\rho)^{-1}CP_{k+i+1|k+i}C^\top \\
&\; \; + \rho^{-1}R]^{-1} (y_{k+i+1} - C(A\hat{x}_{k+i} + Bu_{k+i})) \\
&\ge (1-\beta)(1-\rho)\delta_{k+i}^2 \\
&\ge (1-\beta)^{i+1} (1-\rho)^{i+1} \delta_k^2
\end{align*}
by the fact that $[(1-\rho)^{-1}CP_{k+i+1|k+i} C^\top + \rho^{-1}R]^{-1}$ is positive definite and that $0<\beta<1, \, 0<\rho<1$. Thus, we conclude that $\delta_{k+i}^2 \ge (1-\beta)^i(1-\rho)^i \delta_k^2$ for any $i \ge 0$. From~(\ref{eq.sse_1}), for all $x_{k+i} \in X_{k+i|k+i}$, we have that:
\begin{align*}
&(x_{k+i} - \hat{x}_{k+i})^\top P_{k+i|k+i}^{-1} (x_{k+i} - \hat{x}_{k+i}) \\
\le &1 - \delta_{k+i}^2 \le 1 - (1-\beta)^i(1-\rho)^i \delta_k^2.
\end{align*}
By definition of the estimation error $e_k$, we then have $e_{k+i}^\top P_{k+i|k+i}^{-1} e_{k+i} \le 1 - (1-\beta)^i(1-\rho)^i \delta_k^2$ for any $i \ge 0$, which completes the proof.
\end{proof}

\subsection{Time Invariant Constraint Tightening}
\label{appendix.B}
We focus on the tightened constraint of the form:
\begin{equation}
\label{ineq:cstr-tightening-ss}
F\bar{x}_k + G\bar{u}_k + (F + GK)s - GKe \le f,
\end{equation}
for any $s \in \mathbb{S}_\infty$ and $e \in \tilde{\mathbb{E}}_\infty$, where $\mathbb{S}_\infty$ is defined by~(\ref{eq:s-mRPI}). For brevity, we use $A_K \coloneqq A + BK$ throughout this section.

Let a pair of positive integers $r_1$, $r_2$ and a pair of scalars $0 \le \alpha_1, \alpha_2 <1$ satisfy:
\begin{equation}
\label{appendix.eq-1}
A_K^{r_1} \mathbb{W} \subseteq \alpha_1 \mathbb{W}, \quad A_K^{r_2} BK \tilde{\mathbb{E}}_\infty \subseteq \alpha_2 BK \tilde{\mathbb{E}}_\infty.
\end{equation}
It can be shown that the minimal RPI set approximation for $s_k$ can be determined by~\cite[Chapter~3]{Kouvaritakis2015model}:
\begin{equation*}
\mathbb{S}_\infty \doteq \frac{1}{1-\alpha_1} \bigoplus_{j=0}^{r_1 - 1}A_K^j \mathbb{W} \oplus \frac{1}{1-\alpha_2} \bigoplus_{j=0}^{r_2 - 1} - A_K^j BK\tilde{\mathbb{E}}_\infty.
\end{equation*}
Therefore, the tightened constraint~(\ref{ineq:cstr-tightening-ss}) can be written as:
\begin{align*}
F\bar{x}_k
&+
G\bar{u}_k
+
\frac{1}{1 - \alpha_1} \sum_{j=0}^{r_1 - 1} (F+GK) A_K^jw_j \\
&-
\frac{1}{1 - \alpha_2} \sum_{j=0}^{r_2 - 1} (F+GK) A_K^j BK e_j
-
GKe
\le f,
\end{align*}
for any $w_j \in \mathbb{W}$, $e_j \in \tilde{\mathbb{E}}_\infty$, and $e \in \tilde{\mathbb{E}}_\infty$. The constraint tightening only requires to solve quadratically constrained linear programs under the ellipsoidal $\mathbb{W}$ and $\tilde{\mathbb{E}}_\infty$.

In cases where $\theta \coloneqq BK$ is full row rank, we know that $A_K^{r_2} \theta \tilde{\mathbb{E}}_\infty \subset \alpha_2 \theta \tilde{\mathbb{E}}_\infty$ if and only if $\theta^\dag A_K^{r_2} \theta \tilde{\mathbb{E}}_\infty \subset \alpha_2 \tilde{\mathbb{E}}_\infty$, where $\theta^\dag$ is the Moore-Penrose pseudoinverse of $\theta$. In other words, for all $e \in \tilde{\mathbb{E}}_\infty$, the point $\theta^\dag A_K^{r_2} \theta e$ lies in the scaled ellipsoid $\alpha_2 \tilde{\mathbb{E}}_\infty$. The same reasoning applies to $A_K^{r_1} \mathbb{W} \subset \alpha_1 \mathbb{W}$. Therefore, the condition~(\ref{appendix.eq-1}) is equivalent to:
\begin{align}
\label{appendix.eq-2}
\max_{w \in \mathbb{W}} \, (A_K^{r_1} w)^\top Q^{-1}(A_K^{r_1} w) \le \alpha_1^2, \\
\label{appendix.eq-3}
\max_{e \in \tilde{\mathbb{E}}_\infty} \, (\theta^\dag A_K^{r_2} \, \theta e)^\top P_\infty^{-1}(\theta^\dag A_K^{r_2} \, \theta e) \le \alpha_2^2,
\end{align}
where the maximization is performed elementwise. Given $r_1, \, r_2$, the optimal value of $\alpha_1, \, \alpha_2$ can be computed accordingly. More accurate approximation of the robust positively invariant set can be achieved with larger $r_1, \, r_2$ and smaller $\alpha_1, \, \alpha_2$. Although~(\ref{appendix.eq-2}) and~(\ref{appendix.eq-3}) are non-convex problems, the optimizer~\cite{andersson2019casadi} always returns the optimal solution efficiently with non-zero initializations.

In cases where $\theta$ is row rank-deficient, one can always find a full-dimensional bounding ellipsoid of $\theta \, \tilde{\mathbb{E}}_\infty$ and follow the proposed algorithm. We note that the computation of the constraint tightening described in this section does not require the Minkowski sum, thus making the proposed algorithm generalize well to high-dimensional systems.

\subsection{Proof of Theorem~\ref{theorem-1}}
\label{proof:theorem-1}
\begin{proof}
Assume that at time $k$ the optimal control problem~\cref{prob:Safe-SM-MPC} is feasible and let $\{\bar{x}_{k|k}^*,\bar{x}_{k+1|k}^*,\dots,\bar{x}_{k+N|k}^*\}$ and $\{\bar{u}_{k|k}^*,\bar{u}_{k+1|k}^*,\dots,\bar{u}_{k+N-1|k}^*\}$ be the optimal nominal state trajectory and input sequence respectively. At time $k+1$, we have:
\begin{equation*}
\bar{x}_{k+1}=A\bar{x}_k+B\bar{u}_{k|k}^*=A\bar{x}_{k|k}^*+B\bar{u}_{k|k}^*=\bar{x}_{k+1|k}^*.
\end{equation*}
In addition, by the fact that $\delta_{k+1}^2 \ge (1-\beta)(1-\rho)\delta_k^2$ and Proposition~\ref{prop.1}, we know that for any $i \ge 0$:
\begin{align*}
\mathbb{E}_{k+i+1|k+1} \subseteq \mathbb{E}_{k+i+1|k}, \\
\mathbb{S}_{k+i+1|k+1} \subseteq \mathbb{S}_{k+i+1|k}.
\end{align*}
Thus, the nominal state trajectory
\begin{equation}
\label{seq.s}
\{\bar{x}_{k+1|k}^*,\bar{x}_{k+2|k}^*,\dots,\bar{x}_{k+N|k}^*,(A + BK) \bar{x}_{k+N|k}^*\}
\end{equation}
and the related input sequence
\begin{equation}
\label{seq.c}
\{ \bar{u}_{k+1|k}^*, \bar{u}_{k+2|k}^*, \dots, \bar{u}_{k+N-1|k}^*, K\bar{x}_{k+N|k}^* \}
\end{equation}
is a feasible solution to the problem~\cref{prob:Safe-SM-MPC} at time $k+1$ by Assumption~\ref{assumption-1} and the fact that $\bar{x}_{k+N|k}^* \in \mathbb{X}_k^f$. Therefore, we conclude by induction that the problem~\cref{prob:Safe-SM-MPC} is feasible for any $k > 0$.

The constraint satisfaction follows from the use of the tightened constraint~(\ref{ineq:cstr-tightening-ellipsoids-conservative}) and the fact that the nominal state $\bar{x}_k$ satisfies~(\ref{eq:nominal-system}).

We now show that $V_k^*$ is decreasing along the trajectory. Note that~\cref{seq.s,seq.c} is a suboptimal solution to the problem~(\ref{prob:Safe-SM-MPC}) at time $k+1$, therefore we have:
\begin{align*}
V_{k+1}^* &\le \sum_{i=1}^{N-1} q(\bar{x}_{k+i|k}^*, \bar{u}_{k+i|k}^*) \\
&\quad \quad + q(\bar{x}_{k+N|k}^*, K\bar{x}_{k+N|k}^*) + p((A + BK) \bar{x}_{k+N|k}^*) \\
&= V_k^* - q(\bar{x}_{k|k}^*, \bar{u}_{k|k}^*) - p(\bar{x}_{k+N|k}^*) \\
&\quad \quad + q(\bar{x}_{k+N|k}^*, K\bar{x}_{k+N|k}^*) + p((A + BK) \bar{x}_{k+N|k}^*).
\end{align*}
Therefore we have:
\begin{equation*}
V_{k+1}^* - V_k^* \le - q(\bar{x}_{k|k}^*, \bar{u}_{k|k}^*)
\end{equation*}
by Assumption~\ref{assumption-1} and the fact that $\bar{x}_{k+N|k}^* \in \mathbb{X}_k^f$. Note that the stage cost $q(\cdot,\cdot)$ and the terminal cost $p(\cdot)$ are both positive definite. Therefore, the optimal cost $V_k^*$ is a decreasing Lyapunov function along the closed-loop trajectory, which implies that the nominal state $\bar{x}_k$ converges to the origin as $k \to \infty$. Moreover, we know that $x_k \in \bar{x}_k \oplus \mathbb{S}_k$ by the definition of the control error. Thus, we conclude that $x_k$ converges to $\mathbb{S}_\infty$ as $k \to \infty$.
\end{proof}

\end{document}